\documentstyle [12pt]{article}

\catcode`@=11
\def\makepreprititle{\par
 \begingroup
 \def\thefootnote{\fnsymbol{footnote}}
 \def\@makefnmark{\hbox
 to 0pt{$^{\@thefnmark}$\hss}}
 \if@twocolumn
 \twocolumn[\@makepreprititle]
 \else \newpage
 \global\@topnum\z@ \@makepreprititle \fi\thispagestyle{plain}\@thanks
 \endgroup
 \setcounter{footnote}{0}
 \let\makepreprititle\relax
 \let\@makepreprititle\relax
 \gdef\@preprintnumber{}\gdef\@preprintdate{}\gdef\@thanks{}\gdef\@author{}
 \gdef\@title{}\gdef\@subtitle{}
 \let\thanks\relax}
\def\subtitle#1{\gdef\@subtitle{#1}}
\def\preprintnumber#1{\gdef\@preprintnumber{#1}}
\def\preprintdate#1{\gdef\@preprintdate{#1}}
\def\@makepreprititle{\newpage
 \null
 \begin{flushright} \@preprintnumber \\
 \@preprintdate \end{flushright}
 \vskip 2em \begin{center}
 {\LARGE \@title \par} \vskip 1.5em
 {\large \lineskip .5em
\begin{tabular}[t]{c}\@author
 \end{tabular}\par}
 \vskip 1em {\large \@date} \end{center}
 \par
 \vskip 1.5em}
\def\abstract{\if@twocolumn
\section*{Abstract}
\else \normalsize
\begin{center}
{\bf Abstract\vspace{-.5em}\vspace{0pt}}
\end{center}
\quotation
\fi}
\if@twoside \oddsidemargin 21pt \evensidemargin 59pt
\else \oddsidemargin 5pt \evensidemargin 5pt
\fi
\advance\textheight by \topskip
\def\section{\@startsection {section}{1}{\z@}{3.5ex plus 1ex minus
 .2ex}{2.3ex plus .2ex}{\Large\bf}}
\catcode`@=12

\preprintnumber{{\boldmath $Y\!ukawa$ $Institute$ $Kyoto$}
                 \hspace{72mm}
                 YITP/K-955$\;\;\;\;$}
\preprintdate{November 1991}

\title{Unitary Representations of $W$ Infinity Algebras\vspace{18mm}}
\author{Satoru Odake\thanks{e-mail address: odake@jpnrifp.bitnet.}}
\date   {{\sl Yukawa Institute for Theoretical Physics\\
              Kyoto University, Kyoto 606, Japan}\vspace{8mm}}

\begin{document}

\newcommand {\beq}{\begin{equation}}
\newcommand {\eeq}{\end{equation}}
\newcommand {\beqa}{\begin{eqnarray}}
\newcommand {\eeqa}{\end{eqnarray}}
\newcommand {\beqan}{\begin{eqnarray*}}
\newcommand {\eeqan}{\end{eqnarray*}}
\newcommand {\sfrac}[2]{{\textstyle \frac{#1}{#2}}}
\newcommand {\lfrac}[2]{\frac{\displaystyle #1}{\displaystyle #2}}
\newcommand {\ds}{\displaystyle}
\newcommand {\n}{\nonumber \\}
\newcommand {\cleqn}{\setcounter{equation}{0}}
\newcommand {\eqn}[1]{(\ref{#1})}
\newcommand {\eq}[1]{eq.~(\ref{#1})}
\newcommand {\eqs}[1]{eqs.~(\ref{#1})}
\newcommand {\Eq}[1]{Eq.~(\ref{#1})}
\newcommand {\eql}[1]{eqs.~(\ref{#1},}
\newcommand {\eqr}[1]{\ref{#1})}
\newcommand {\eqm}[1]{\ref{#1},}
\newcommand {\scs}{\scriptsize}
\newcommand {\Label}[1]{\label{#1}}
\newcommand {\Bibitem}[1]{\bibitem{#1}}
\newcommand {\eqdef}{\stackrel{\rm def}{=}}
%
%
%
\font\tennn=msym10                                          
\font\twelvenn=msym10 scaled\magstep1                       
%
%
%
\newcommand {\Sym}[1]{\leavevmode\raise-.25ex\hbox{\twelvenn #1}}
\newcommand {\sym}[1]{\leavevmode\raise-.15ex\hbox{\tennn #1}}
\newcommand {\SymR}{\mbox{\twelvenn R}}
\newcommand {\symR}{\mbox{\tennn R}}

\makepreprititle
\begin{abstract}
We study the irreducible unitary highest weight representations,
which are obtained from free field realizations,
of $W$ infinity algebras
($W_{\infty}$, $W_{1+\infty}$, $W_{\infty}^{1,1}$,
$W_{\infty}^M$, $W_{1+\infty}^N$, $W_{\infty}^{M,N}$)
with central charges ($2$, $1$, $3$, $2M$, $N$, $2M+N$).
The characters of these representations are computed.

We construct a new extended superalgebra $W_{\infty}^{M,N}$,
whose bosonic sector is $W_{\infty}^M\oplus W_{1+\infty}^N$.
Its representations obtained from a free field realization with
central charge $2M+N$, are classified into two classes: continuous
series and discrete series. For the former there exists a supersymmetry,
but for the latter a supersymmetry exists only for $M=N$.
\end{abstract}
\newpage
%
%
\section {Introduction}
The conformal field theory (CFT) in two dimensional space-time
has made great progress in close contact with both the string theory
and various branches of mathematics. The Virasoro algebra plays a
central role in CFT, and to construct models of CFT and extend this
theory, one needs an extension of the Virasoro algebra. In view of
this several extensions (superconformal algebras, $W$ algebras,
parafermions, etc.) have been studied. The notable example is
Zamolodchikov's $W_3$ algebra and its $W_N$ generalization
($A_{N-1}$ type $W$ algebra) containing fields of conformal
weight (spin) $2,\cdots,N$ as conserved currents \cite{ZFL}.
All of extended Virasoro algebras containing currents of spin $>2$
have a non-linear property.

By taking an appropriate $N\rightarrow\infty$ limit of the $W_N$ algebra,
one can obtain a linear algebra with infinite number of fields.
The first example is the $w_{\infty}$ algebra \cite{Bakas}, which can be
interpreted as the algebra of area-preserving diffeomorphisms
of two dimensional phase space. But $w_{\infty}$ admits a central
extension only in the Virasoro sector. By deforming $w_{\infty}$,
Pope, Romans and Shen constructed the $W_{\infty}$ algebra \cite{PRS},
which admits central extension in all spin sectors. This is another
large $N$ limit of $W_N$. In addition, they constructed the
$W_{1+\infty}$ algebra \cite{PRS2}, which contains a spin 1 field too,
and their super extension, super $W_\infty$ algebra \cite{sW}, whose
bosonic sector is $W_{\infty}\oplus W_{1+\infty}$. Soon afterward
Bakas and Kiritsis constructed the $W_{\infty}^M$ algebra
\cite{BK}, which is a $u(M)$ matrix version of $W_{\infty}$,
and Sano and the present author constructed
the $\widehat{su}(N)$-$W_{1+\infty}$ algebra \cite{OS}, which
is an extension of $W_{1+\infty}$ and contains the $SU(N)$ current
algebra. We will change the notation $\widehat{su}(N)$-$W_{1+\infty}$
to $W_{1+\infty}^N$.
$W_{1+\infty}^N$ is to $W_{\infty}^N$ what $W_{1+\infty}$ is to
$W_{\infty}$ \cite{OS2}. In this paper we will construct a new
superalgebra $W_{\infty}^{M,N}$, whose bosonic sector is
$W_{\infty}^M\oplus W_{1+\infty}^N$. In this notation, super
$W_{\infty}$ \cite{sW} and super $\widehat{su}(N)$-$W_{\infty}$
\cite{OS} are $W_{\infty}^{1,1}$ and $W_{\infty}^{1,N}$ respectively.

Given an algebra which generates the symmetry of a model, it is
important to develop its representation theory to find
what fields appear in the model. The success of the
representation theory of the Virasoro algebra is a good example
\cite{FQS}. In spite of difficulties due to the non-linearity,
the minimal representations of the $W_N$ algebra have been studied
in detail by using the free field realization of Feigin-Fuchs
type or the coset model of affine Lie algebras \cite{ZFL,BBSS}.
Although $W$ infinity algebras are linear algebras and
their structure constants are explicitly known,
their representation theories have been studied very poorly
\cite{BK3,OS2}. In this paper we will initiate the study of
representation theories of $W$ infinity algebras and construct their
unitary representations based on free field realizations.

Our methods to study representation theories are as follows.
First we prepare a free field realization of the $W$ infinity algebra
and its Fock space. In the Fock space, we try finding all the highest
weight states (HWS's) of the subalgebra (an affine Lie algebra or
the Virasoro algebra), whose generators have the lowest spin.
Next we check these states are also the HWS's of the whole $W$ infinity
algebra. To compute the characters of $W_{1+\infty}$ and
$W_{1+\infty}^N$, we use the character formulas of the affine Lie
algebras and the fact that the generators of $W_{1+\infty}$ and
$W_{1+\infty}^N$ do not change the $U(1)$ charge. For $W_{\infty}$ and
$W_{\infty}^M$, we use the facts that the generators of the $W$
infinity algebras are represented in terms of bilinears of the free
fields and they preserve a certain quantum number.
For super cases, we use the results of bosonic cases and the spectral
flow invariance. The Witten index is also computed.

The organization of this paper is as follows. In \S 2 we construct
a new superalgebra $W_{\infty}^{M,N}$ and present its free field
realization. In \S 3-5 the representations of $W$ infinity algebras
($W_{1+\infty}$, $W_{1+\infty}^N$, $W_{\infty}$, $W_{\infty}^M$,
$W_{\infty}^{1,1}$, $W_{\infty}^{M,N}$) are studied, and their
characters are computed. In \S 6 we present a discussion.

\section {Algebras and Free Field Realizations}

We will define a new extended superalgebra $W_{\infty}^{M,N}$,
whose bosonic sector is $W_{\infty}^M\oplus W_{1+\infty}^N$.
Other $W$ infinity algebras are contained in it as subalgebras.
$W_{\infty}^{M,N}$ is generated by\footnote{
 As usual, the mode expansion of a field $A(z)$ of conformal weight $h$
 is $A(z)=\sum A_nz^{-n-h}$, where the sum is taken over
 $n\in\Sym{Z}-h$ for the Neveu-Schwarz ($NS$) sector,
 $n\in\Sym{Z}$ for the Ramond ($R$) sector.}
\beqa
  W^{i,(\alpha\beta)}(z),&&(i\geq -1;\alpha,\beta=1,2,\cdots,N), \\
  \tilde{W}^{i,(ab)}(z),&&(i\geq 0;a,b=1,2,\cdots,M), \\
  G^{i,a\alpha}(z),&&(i\geq 0;a=1,\cdots,M;\alpha=1,\cdots,N), \\
  \bar{G}^{i,a\alpha}(z),&&(i\geq 0;a=1,\cdots,M;\alpha=1,\cdots,N).
\eeqa
$W^{i,(\alpha\beta)}(z)$ and $\tilde{W}^{i,(ab)}(z)$ are bosonic
fields with conformal spin $i+2$, and generate $W_{1+\infty}^N$ and
$W_{\infty}^M$ respectively.
$W_{1+\infty}^N$ ($W_{\infty}^M$) contains $W_{1+\infty}$
($W_{\infty}$) as a subalgebra, and their generators are
\beq
  V^i(z)=\sum_{\alpha=1}^N W^{i,(\alpha\alpha)}(z),~~
  \tilde{V}^i(z)=\sum_{a=1}^M \tilde{W}^{i,(aa)}(z).
\eeq
$V^0(z)$ and $\tilde{V}^0(z)$ are the Virasoro generators with
central charge $c$ and $\tilde{c}$ respectively.
$W_{1+\infty}^N$ contains the $U(N)$ current algebra, which is
generated by
\beqa
   \widehat{su}(N)_k
   &:&
   \left \{
   \begin{array}{l}
     H^i(z)=J^{(ii)}(z)-J^{(i+1,i+1)}(z),~~(i=1,\cdots,N-1)~~
     \mbox{Cartan} \\
     J^{(\alpha\beta)}(z),~~(\alpha<\beta)~~\mbox{raising };~~
     (\alpha>\beta)~~\mbox{lowering},
   \end{array}
   \right.
   \\
   \hat{u}(1)_K
   &:&
   J(z)=\sum_{\alpha=1}^N J^{(\alpha\alpha)}(z),
\eeqa
where $J^{(\alpha\beta)}(z)=-4qW^{-1,(\beta\alpha)}(z)$
\footnote{This definition is slightly different from \cite{OS}.}.
$k$ is the level of $\widehat{su}(N)$ and $K$ stands for a normalization
of $\hat{u}(1)$ ($\lbrack J_m,J_n \rbrack =Km\delta_{m+n,0}$).
$G^{i,a\alpha}(z)$ and $\bar{G}^{i,a\alpha}(z)$ are fermionic
fields with conformal spin $i+\frac{3}{2}$.
$W^{i,(\alpha\beta)}(z)$ ($\tilde{W}^{i,(ab)}(z)$,
$G^{i,a\alpha}(z)$, $\bar{G}^{i,a\alpha}(z)$) transform
according to the adjoint (trivial, ${\bf N}$, ${\bf \bar{N}}$)
representation of $su(N)$, and have $U(1)$ charge $0$ ($0,1,-1$),
respectively.

(Anti-)commutation relations of $W_{\infty}^{M,N}$ are given
by\footnote{
 $q$ is a deformation parameter and we can take it to be an arbitrary
 non-zero constant(e.g., $q=\frac{1}{4}$).
 This $q$ has nothing to do with $q=e^{2\pi i\tau}$ appearing in
 the characters.}
\beqa
   \lbrack W^{i,(\alpha\beta)}_m,W^{j,(\gamma\delta)}_n\rbrack &\!\!=\!\!&
       \sfrac{1}{2}\sum_{r\geq -1}
        q^r g^{ij}_r(m,n)
        (\delta^{\gamma\beta}W^{i+j-r,(\alpha\delta)}_{m+n}
         +(-1)^r\delta^{\alpha\delta}W^{i+j-r,(\gamma\beta)}_{m+n})  \n
    &&  +\delta^{ij}\delta^{\alpha\delta}
         \delta^{\gamma\beta}\delta_{m+n,0}q^{2i}k_i(m), \\
   \lbrack \tilde{W}^{i,(ab)}_m,\tilde{W}^{j,(cd)}_n\rbrack &\!\!=\!\!&
       \sfrac{1}{2}\sum_{r\geq -1}
        q^r \tilde{g}^{ij}_r(m,n)
        (\delta^{cb}\tilde{W}^{i+j-r,(ad)}_{m+n}
         +(-1)^r\delta^{ad}\tilde{W}^{i+j-r,(cb)}_{m+n})  \n
    &&  +\delta^{ij}\delta^{ad}
         \delta^{cb}\delta_{m+n,0}q^{2i}\tilde{k}_i(m), \\
   \lbrack W^{i,(\alpha\beta)}_m,G^{j,c\gamma}_n\rbrack &\!\!=\!\!&
       \delta^{\alpha\gamma}\sum_{r\geq -1}
        q^r a^{ij}_r(m,n)G^{i+j-r,c\beta}_{m+n}, \\
   \lbrack W^{i,(\alpha\beta)}_m,\bar{G}^{j,c\gamma}_n\rbrack &\!\!=\!\!&
       \delta^{\beta\gamma}\sum_{r\geq -1}
        q^r (-1)^ra^{ij}_r(m,n)\bar{G}^{i+j-r,c\alpha}_{m+n}, \\
   \lbrack \tilde{W}^{i,(ab)}_m,G^{j,c\gamma}_n\rbrack &\!\!=\!\!&
       \delta^{bc}\sum_{r\geq -1}
        q^r \tilde{a}^{ij}_r(m,n)G^{i+j-r,a\gamma}_{m+n}, \\
   \lbrack \tilde{W}^{i,(ab)}_m,\bar{G}^{j,c\gamma}_n\rbrack &\!\!=\!\!&
       \delta^{ac}\sum_{r\geq -1}
        q^r (-1)^r\tilde{a}^{ij}_r(m,n)\bar{G}^{i+j-r,b\gamma}_{m+n}, \\
   \lbrace G^{i,a\alpha}_m,\bar{G}^{j,b\beta}_n\rbrace &\!\!=\!\!&
       \sum_{r\geq 0}
        q^r (\delta^{ab}b^{ij}_r(m,n)W^{i+j-r,(\beta\alpha)}_{m+n}
            +\delta^{\alpha\beta}\tilde{b}^{ij}_r(m,n)
                                 \tilde{W}^{i+j-r,(ab)}_{m+n})   \n
    && +\delta^{ij}\delta^{ab}\delta^{\alpha\beta}\delta_{m+n,0}
                   q^{2i}\check{k}_i(m), \\
   \lbrack W^{i,(\alpha\beta)}_m,\tilde{W}^{j,(ab)}_n\rbrack &\!\!=\!\!&
     \lbrace G^{i,a\alpha}_m,G^{j,b\beta}_n\rbrace ~=~
     \lbrace \bar{G}^{i,a\alpha}_m,\bar{G}^{j,b\beta}_n\rbrace ~=~0.
\eeqa
The structure constants \cite{PRS,PRS2,sW,BK,OS} are
\beqa
   g^{ij}_r(m,n)&\!\!=\!\!&
      \sfrac{1}{2(r+1)!}\phi^{ij}_r(0,-\sfrac{1}{2})N^{i,j}_r(m,n), \\
   \tilde{g}^{ij}_r(m,n)&\!\!=\!\!&
      \sfrac{1}{2(r+1)!}\phi^{ij}_r(0,0)N^{i,j}_r(m,n), \\
   a^{ij}_r(m,n)&\!\!=\!\!&
      \sfrac{(-1)^r}{4(r+2)!}
      ((i+1)\phi^{ij}_{r+1}(0,0)-(i-r-1)\phi^{ij}_{r+1}(0,-\sfrac{1}{2}))
      N^{i,j-\frac{1}{2}}_r(m,n), \\
   \tilde{a}^{ij}_r(m,n)&\!\!=\!\!&
      \sfrac{-1}{4(r+2)!}
      ((i-r)\phi^{ij}_{r+1}(0,0)-(i+2)\phi^{ij}_{r+1}(0,-\sfrac{1}{2}))
      N^{i,j-\frac{1}{2}}_r(m,n), \\
   b^{ij}_r(m,n)&\!\!=\!\!&
      \sfrac{(-1)^r4}{r!}
      ((i+j+2-r)\phi^{ij}_r(\sfrac{1}{2},-\sfrac{1}{4})  \n
    && ~~~~~~~~~
      -(i+j+\sfrac{3}{2}-r)\phi^{ij}_{r+1}(\sfrac{1}{2},-\sfrac{1}{4}))
      N^{i-\frac{1}{2},j-\frac{1}{2}}_{r-1}(m,n), \\
   \tilde{b}^{ij}_r(m,n)&\!\!=\!\!&
      -\sfrac{4}{r!}
      ((i+j+1-r)\phi^{ij}_r(\sfrac{1}{2},-\sfrac{1}{4})  \n
    && ~~~~~~~~~
      -(i+j+\sfrac{3}{2}-r)\phi^{ij}_{r+1}(\sfrac{1}{2},-\sfrac{1}{4}))
      N^{i-\frac{1}{2},j-\frac{1}{2}}_{r-1}(m,n),
\eeqa
and
\beqa
   \hspace{-20mm}
   &&N^{x,y}_r(m,n)=
     \sum_{\ell=0}^{r+1}(-1)^{\ell} {r+1\choose \ell}
     \lbrack x+1+m \rbrack_{r+1-\ell}
     \lbrack x+1-m \rbrack_{\ell} \n
   \hspace{-20mm}
   &&\hspace{53mm}
     \times
     \lbrack y+1+n \rbrack_{\ell}
     \lbrack y+1-n \rbrack_{r+1-\ell}, \\
   \hspace{-20mm}
   &&\phi^{ij}_r(x,y)=\!
     ~_4F_3 \Bigl[
       \left.
       \begin{array}{c}
         -\frac{1}{2}-x-2y,\frac{3}{2}-x+2y,
         -\frac{r+1}{2}+x,-\frac{r}{2}+x \\
         -i-\frac{1}{2},-j-\frac{1}{2},i+j-r+\frac{5}{2}  \\
       \end{array}
       \right.
       ;1 \Bigr],\\
   \hspace{-20mm}
   &&~_4F_3 \Bigl[
       \left.
       \begin{array}{c}
         a_1,a_2,a_3,a_4 \\
         b_1,b_2,b_3  \\
       \end{array}
       \right.
       ;z \Bigr]
    =\sum_{n=0}^{\infty}
        \frac{(a_1)_n(a_2)_n(a_3)_n(a_4)_n}
             {(b_1)_n(b_2)_n(b_3)_n}
        \frac{z^n}{n!},
\eeqa
where $[x]_n=x(x-1)\cdots(x-n+1)$, $[x]_0=1$ and
$(x)_n=x(x+1)\cdots(x+n-1)$, $(x)_0=1$ and ${x \choose n}=[x]_n/n!$.
Since $g^{ij}_r=b^{ij}_r=0$ for $i-j-r<-1$ and
$\tilde{g}^{ij}_r=a^{ij}_r=\tilde{a}^{ij}_r=\tilde{b}^{ij}_r=0$
for $i-j-r<0$, the summations over $r$ are finite sums and
the algebra closes.
The central terms are
\beqa
   k_i(m)=k_i\prod_{j=-i-1}^{i+1}(m+j)~, \hspace{8mm}&~~~&
   k_i=\frac{2^{2i-2}((i+1)!)^2}{(2i+1)!!(2i+3)!!}k, \\
   \tilde{k}_i(m)=\tilde{k}_i\prod_{j=-i-1}^{i+1}(m+j)~, \hspace{8mm}&~~~&
   \tilde{k}_i=\frac{2^{2i-3}i!(i+2)!}{(2i+1)!!(2i+3)!!}\tilde{k}, \\
   \check{k}_i(m)=\check{k}_i\prod_{j=-i-1}^i(m+j+\frac{1}{2})~, &~~~&
   \check{k}_i=\frac{2^{2i}i!(i+1)!}{3((2i+1)!!)^2}\check{k}.
\eeqa
In the case of $W_{\infty}^{M,N}$, the Jacobi identity requires
\beq
   K=Nk~,
   \hspace{5mm}c=Nk~,
   \hspace{5mm}\tilde{c}=M\tilde{k}~,
   \hspace{5mm}\tilde{k}=2k~,
   \hspace{5mm}\check{k}=3k.
\eeq
Since the level $k$ of $\widehat{su}(N)$ $(N>1)$ is a positive integer
for unitary representations, central charges $c$ and $\tilde{c}$ must
be multiples of $N$ and $2M$ respectively.
(Anti-)commutation relations of $W_{\infty}^{M,N}$ are consistent
with the hermiticity properties of the generators:
\beq
   W^{i,(\alpha\beta)\dagger}_n= W^{i,(\beta\alpha)}_{-n},~
   \tilde{W}^{i,(ab)\dagger}_n= \tilde{W}^{i,(ba)}_{-n},~
   G^{i,a\alpha\dagger}_n= \bar{G}^{i,a\alpha}_{-n}.
   \Label{h}
\eeq

The Cartan subalgebra of $W_{\infty}^{M,N}$ is generated by
\beq
  W^{i,(\alpha\alpha)}_0,~~\tilde{W}^{i,(aa)}_0.
\eeq
The HWS of $W_{\infty}^{M,N}$ in the $NS$ sector is defined by
\beq
  \left \{
  \begin{array}{ll}
    A_n|\mbox{hws}\rangle=0, & (n>0;A=W,\tilde{W},G,\bar{G}) \\
    W^{i,(\alpha\beta)}_0|\mbox{hws}\rangle=0, & (\alpha>\beta) \\
    \tilde{W}^{i,(ab)}_0|\mbox{hws}\rangle=0, & (a>b),
  \end{array}
  \right.
\eeq
and, in the $R$ sector, we require one more condition:
\beq
    G^{i,a\alpha,R}_0|\mbox{hws}\rangle^R=0.
\eeq
The HWS's of other $W$ infinity algebras are defined in a similar way.

Since $W_{\infty}^{M,N}$ contains a current algebra,
there exists an automorphism, so called spectral flow \cite{SS}.
Namely (anti-)commutation relations are invariant under the
transformations of the generators. Explicit forms of the transformation
rules are essentially the same as $W_{\infty}^{1,N}$ \cite{OS}.
Due to this property, representations in the $R$ sector and those in
the $NS$ sector have one-to-one correspondence. We define the
representations in the $R$ sector as those mapped from the $NS$ sector
by the spectral flow with $\eta=\frac{1}{2}$.
Then we can show $|\mbox{hws}\rangle^R=|\mbox{hws}\rangle^{NS}$,
because
\beq
  \left \{
  \begin{array}{ll}
    W^{i\prime}_n=\sum_{j=-1}^i\mbox{(coeff.)}\cdot W^j_n
                  +\mbox{(coeff.)}\cdot \delta_{n0}, &
    G^{i\prime}_n=\sum_{j=0}^i\mbox{(coeff.)}\cdot G^j_{n+\frac{1}{2}} \\
    \tilde{W}^{i\prime}_n=\sum_{j=0}^i\mbox{(coeff.)}\cdot \tilde{W}^j_n
                  +\mbox{(coeff.)}\cdot \delta_{n0}, &
    \bar{G}^{i\prime}_n=\sum_{j=0}^i
                  \mbox{(coeff.)}\cdot \bar{G}^j_{n-\frac{1}{2}}.
  \end{array}
  \right.
\eeq

$W_{\infty}^{M,N}$ with level $k\!=\!1$ is realized by $N$ complex free
fermions
$\psi^{\alpha}(z)$ $=$ $\sum_n\psi^{\alpha}_nz^{-n-\frac{1}{2}}$
$(\alpha=1,\cdots,N)$ and $M$ complex free bosons
$i\partial\varphi^a(z)$ $=$ $\sum_n\alpha^a_nz^{-n-1}$
$(a=1,\cdots,M)$.
Operator product expansions of the free fields are
\beq
   \bar{\psi}^{\alpha}(z)\psi^{\beta}(w)
   \sim \frac{\delta^{\alpha\beta}}{z-w},~~~
   i\partial\bar{\varphi}^a(z)i\partial\varphi^b(w)
   \sim \frac{\delta^{ab}}{(z-w)^2}.
   \Label{fope}
\eeq
Generators of $W_{\infty}^{M,N}$ are represented in terms of bilinears
of the free fields \cite{PRS2,sW,BK,OS} :
\beqa
   W^{j,(\alpha\beta)}(z)&\!=\!&\frac{2^{j-1}(j+1)!}{(2j+1)!!}q^j
     \sum_{r=0}^{j+1}(-1)^r {j+1 \choose r}^2
     (\partial^{j+1-r}\bar{\psi}^{\alpha}
      \partial^r\psi^{\beta})(z), \\
   \tilde{W}^{j,(ab)}(z)&\!=\!&\frac{2^{j-1}(j+2)!}{(2j+1)!!}q^j
     \sum_{r=0}^j\frac{(-1)^r}{j+1}
     {j+1 \choose r}{j+1 \choose r+1}
     (\partial^{j-r}i\partial\bar{\varphi}^a
      \partial^ri\partial\varphi^b)(z), \\
   G^{j,a\alpha}(z)&\!=\!&\frac{2^{j+\frac{1}{2}}(j+1)!}{(2j+1)!!}q^j
     \sum_{r=0}^j(-1)^r {j+1 \choose r}{j \choose r}
     (\partial^{j-r}i\partial\bar{\varphi}^a
      \partial^r\psi^{\alpha})(z), \\
   \bar{G}^{j,a\alpha}(z)&\!=\!&\frac{2^{j+\frac{1}{2}}(j+1)!}{(2j+1)!!}q^j
     \sum_{r=0}^j(-1)^{j+r} {j+1 \choose r}{j \choose r}
     (\partial^{j-r}i\partial\varphi^a
      \partial^r\bar{\psi}^{\alpha})(z),
\eeqa
where the normal ordered product of two fields $A(z)$ and $B(z)$ is
defined by $(AB)(z)=\oint_z\frac{dx}{2\pi i}\frac{1}{x-z}A(x)B(z)$.
One of the methods to obtain the general level $k$ realization is
to prepare $k$ copies of the above realization, because
$W_{\infty}^{M,N}$ is linear.
The spectral flow transformation rules of the generators with a
parameter $\eta$ are easily derived from those of the free fields:
\beq
   \psi^{\alpha\prime}(z)=z^{\eta}\psi^{\alpha}(z),~
   \bar{\psi}^{\alpha\prime}(z)=z^{-\eta}\bar{\psi}^{\alpha}(z),~
   \varphi^{a\prime}(z)=\varphi^a(z),~
   \bar{\varphi}^{a\prime}(z)=\bar{\varphi}^a(z),
   \Label{frt}
\eeq
because \eq{fope} are invariant under this transformation.
The transformation rules that will be needed later are
\beq
  \begin{array}{rcl}
    W^{-1,(\alpha\beta)\prime}_n &\!\!=\!\!&
     W^{-1,(\alpha\beta)}_n
     -\delta^{\alpha\beta}\delta_{n0}\frac{1}{4q}\eta, \\
    W^{0,(\alpha\beta)\prime}_n &\!\!=\!\!&
     W^{0,(\alpha\beta)}_n
     -4q\eta W^{-1,(\alpha\beta)}_n
     +\delta^{\alpha\beta}\delta_{n0}\frac{1}{2}\eta^2, \\
    G^{0,a\alpha\prime}_n &\!\!=\!\!&
     G^{0,a\alpha}_{n+\eta},~~~
    \bar{G}^{0,a\alpha\prime}_n =
     \bar{G}^{0,a\alpha}_{n-\eta},~~~
    \tilde{W}^{0,(ab)\prime}_n =
     \tilde{W}^{0,(ab)}_n.
  \end{array}
\eeq

The vacuum states of the fermion and boson Fock spaces,
$|0\rangle$ and $|\vec{p},\vec{\bar{p}}\rangle$, are defined as usual:
$\psi^{\alpha}_m|0\rangle$ $=$ $\bar{\psi}^{\alpha}_n|0\rangle$ $=$
$\alpha^a_n|\vec{p},\vec{\bar{p}}\rangle$ $=$
$\bar{\alpha}^a_n|\vec{p},\vec{\bar{p}}\rangle=0$, $(m\geq 0,n>0)$,
$\alpha^a_0|\vec{p},\vec{\bar{p}}\rangle$ $=$
$p_a|\vec{p},\vec{\bar{p}}\rangle$,
$\bar{\alpha}^a_0|\vec{p},\vec{\bar{p}}\rangle$ $=$
$\bar{p}_a|\vec{p},\vec{\bar{p}}\rangle$.
Hermiticity properties of the generators \eq{h} are satisfied
by those of the free fields
($\psi^{\alpha\dagger}_n=\bar{\psi}^{\alpha}_{-n}$,
$\alpha^{a\dagger}_n=\bar{\alpha}^a_{-n}$).
In the following we take $p_a^*=\bar{p}_a$, so that the unitarity
of the representations is manifest.

\section {Representations of $W_{1+\infty}$ and $W_{1+\infty}^N$}

We first consider the representations of $W_{1+\infty}$ with $c=1$
realized by one complex free fermion. We remark that the Virasoro
generator $V^0(z)$ agrees with the Sugawara form of $\hat{u}(1)$
current. For each integer $n$, we can find the HWS of the
subalgebra $\hat{u}(1)$ contained in the fermion Fock space, and
we denote them as
\beq
  |n\rangle \eqdef
  \left \{
  \begin{array}{ll}
    \psi_{-\frac{1}{2}}\psi_{-\frac{3}{2}}
    \cdots \psi_{-n+\frac{1}{2}} |0\rangle & n\geq 1\\
    |0\rangle & n=0  \\
    \bar{\psi}_{-\frac{1}{2}}\bar{\psi}_{-\frac{3}{2}}
    \cdots \bar{\psi}_{n+\frac{1}{2}} |0\rangle & n\leq -1.
  \end{array}
  \right.
  \Label{hwsW1inf}
\eeq
These states are well known in Sato theory \cite{DJKM}.
We can check that $|n\rangle$ is not only the HWS
of $\hat{u}(1)$ but also the HWS of $W_{1+\infty}$.
The conformal weight $h_n$ and $U(1)$ charge $Q_n$ of $|n\rangle$ are
\beq
  h_n=\frac{1}{2}n^2,~~Q_n=n.
\eeq
Although the eigenvalues of the higher-spin generators are easily
calculated, we omit them here.

Since the dependence on the eigenvalues of higher-spin generators are
very complicated, we consider the characters which count conformal
weight and $U(1)$ charge only:
\beq
  ch^{W_{1+\infty}}(\theta,\tau)
  \eqdef
  \mbox{tr}q^{V^0_0-\frac{1}{24}}z^{J_0},
\eeq
where $q=e^{2\pi i\tau}$ $(\mbox{Im}\tau>0)$ and $z=e^{i\theta}$.
Since $W_{1+\infty}$ contains $\hat{u}(1)$ as a subalgebra, the
representation of $W_{1+\infty}$ has more states than one of
$\hat{u}(1)$. On the other hand, the representation of
$W_{1+\infty}$ has less states than the Fock space with the fixed
$U(1)$ charge, because generators of $W_{1+\infty}$ do not change
$U(1)$ charge. These statements are expressed in terms of characters as
follows:
\beq
  \chi^{\hat{u}(1)_1}_n(\theta,\tau) \leq
  ch^{W_{1+\infty}}_n(\theta,\tau) \leq
  z^n\chi^{Fock}_n(\tau),
\eeq
where $A\leq B$ means $B-A$ is a $q$-series with non-negative
coefficients. In general, the character formula of $\hat{u}(1)_K$ with $U(1)$
charge $Q$ is
\beq
  \chi^{\hat{u}(1)_K}_Q(\theta,\tau) \eqdef
  \mbox{tr}q^{L_0-\frac{1}{24}}z^{J_0} =
  \frac{1}{\eta(\tau)} q^{\frac{1}{2K}Q^2}z^Q,
  \Label{chu1}
\eeq
where $\eta(\tau)=q^{\frac{1}{24}}\prod_{n=1}^{\infty}(1-q^n)$.
On the other hands, the generating function of $\chi^{Fock}_n(\tau)$ is
\beq
  \sum_{n\in\sym{Z}}z^n\chi^{Fock}_n(\tau)
  =
  \mbox{tr}_{Fock} q^{V^0_0-\frac{1}{24}}z^{J_0}
  =
  q^{-\frac{1}{24}} \prod_{n=1}^{\infty}
    (1+zq^{n-\frac{1}{2}})(1+z^{-1}q^{n-\frac{1}{2}}).
\eeq
Due to the Jacobi's triple product identity, we have
$\chi^{\hat{u}(1)_1}_n(\theta,\tau)=z^n\chi^{Fock}_n(\tau)$.
Therefore we obtain the character formula of $W_{1+\infty}$,
\beq
  ch^{W_{1+\infty}}_n(\theta,\tau) =
  \chi^{\hat{u}(1)_1}_n(\theta,\tau).
\eeq

Next we consider the representations of $W_{1+\infty}^N$ with $c=N$
realized by $N$ complex free fermions. We use the same techniques as
$W_{1+\infty}$ case. We remark that, in the case of level $k\!=\!1$,
the Virasoro generator $V^0(z)$ agrees with the sum of the Sugawara
form of $\hat{u}(1)_N$ and $\widehat{su}(N)_1$ \cite{GNO}.
For each integer $n$, there exists the HWS of
$\widehat{su}(N)_1$, and we denote them as
\beq
  |n\rangle \eqdef
  \left \{
  \begin{array}{ll}
    \prod_{j=1}^m
     (\prod_{\alpha=1}^N \psi^{\alpha}_{-j+\frac{1}{2}})
    \cdot
    \prod_{\alpha=1}^a \psi^{\alpha}_{-m-\frac{1}{2}}
    |0\rangle & n\geq 1\\
    |0\rangle & n=0\\
    \prod_{j=1}^{-m-1}
     (\prod_{\alpha=1}^N \bar{\psi}^{N+1-\alpha}_{-j+\frac{1}{2}})
    \cdot
    \prod_{\alpha=1}^{N-a} \bar{\psi}^{N+1-\alpha}_{m+\frac{1}{2}}
    |0\rangle & n\leq -1,
  \end{array}
  \right.
  \Label{hwsW1infN}
\eeq
where we express $n$ as $n=Nm+a$, $(m\in\Sym{Z};a=0,1,\cdots,N-1)$.
$|n\rangle$ is the HWS of the $a$-th rank antisymmetric representation
of $\widehat{su}(N)_1$.
The state $|n\rangle$ is also the HWS of $W_{1+\infty}^N$.
The conformal weight $h_n$ and $U(1)$ charge $Q_n$ are
\beq
  h_n=\frac{1}{2N}n^2+\frac{a(N-a)}{2N},~~Q_n=n.
\eeq
The first and second factors of $h_n$ are contributions from
$\hat{u}(1)_N$ and $\widehat{su}(N)_1$ respectively.

Neglecting the dependence on the eigenvalues of higher-spin generators,
we consider the characters which count conformal weight, $U(1)$ charge
and eigenvalues of $SU(N)$,
\beq
  ch^{W_{1+\infty}^N}(\theta,\vec{\theta},\tau)
  \eqdef
  \mbox{tr}q^{V^0_0-\frac{N}{24}}e^{i\theta J_0}
           e^{i\vec{\theta}\cdot\vec{H}_0},
\eeq
where $\vec{\theta}=\sum_{i=1}^{N-1}\theta_i\vec{\alpha}_i$
and $\vec{\alpha}_i$ is a simple root of $su(N)$.
$W_{1+\infty}^N$ contains $\hat{u}(1)_N\oplus\widehat{su}(N)_1$ as a
subalgebra, and generators of $W_{1+\infty}^N$ do not change $U(1)$
charge. Therefore, the similar argument as $W_{1+\infty}$ case
shows that the character formula of $W_{1+\infty}^N$ is given by
\beq
  ch^{W_{1+\infty}^N}_n(\theta,\vec{\theta},\tau)
  =
  \chi^{\hat{u}(1)_N}_n(\theta,\tau)
  \chi^{\widehat{su}(N)_1}_a(\vec{\theta},\tau),
  \Label{chW1infN}
\eeq
where $a\equiv n(\mbox{mod } N),0\leq a\leq N-1$.
The character of $\hat{u}(1)$ is given by \eq{chu1}, and the character
formula of $\widehat{su}(N)_1$ is given by \cite{GO}
\beq
  \chi^{\widehat{su}(N)_1}_a(\vec{\theta},\tau)
  \eqdef
  \mbox{tr}q^{L_0-\frac{N-1}{24}}e^{i\vec{\theta}\cdot\vec{H}_0}
  =
  \frac{1}{\eta(\tau)^{N-1}}
  \sum_{\vec{M} \in \Lambda_R}
     q^{\frac{1}{2}(\vec{M}+\vec{\Lambda}_a)^2}
     e^{i\vec{\theta}\cdot (\vec{M}+\vec{\Lambda}_a)},
  \Label{chsuN}
\eeq
where $\Lambda_R$ is a root lattice of $su(N)$, and
$\vec{\Lambda}_i$ $(1\leq i\leq N-1)$ is a fundamental weight of
$su(N)$ and $\vec{\Lambda}_0=\vec{0}$.
To show \eq{chW1infN}, we need the identity:
\beqa
  \mbox{tr}_{Fock}
     q^{V^0_0-\frac{N}{24}}e^{i\theta J_0}
     e^{i\vec{\theta}\cdot\vec{H}_0}
  &=&
  q^{-\frac{N}{24}}
  \prod_{j=1}^N\prod_{n=1}^{\infty}
    (1+zz_{j-1}^{-1}z_jq^{n-\frac{1}{2}})
    (1+z^{-1}z_{j-1}z_j^{-1}q^{n-\frac{1}{2}}) \n
  &=&
  \sum_{n \in \sym{Z}}
    \chi^{\hat{u}(1)_N}_n(\theta,\tau)
    \chi^{\widehat{su}(N)_1}_a(\vec{\theta},\tau),
\eeqa
where $z_j=e^{i\theta_j}$, $z_0=z_N=1$, and $a\equiv n (\mbox{mod } N)$.
This identity is proved by the Jacobi's triple product identity.

\section {Representations of $W_{\infty}$ and $W_{\infty}^M$}

We first consider the representations of $W_{\infty}$ with
$\tilde{c}=2$ realized by one complex free boson. In the boson Fock
space, the HWS's of the Virasoro generator $\tilde{V}^0(z)$ are
classified into two classes: continuous series, whose momentum can be
changed continuously, and discrete series, whose state exists for each
integer $n$,
\beqa
  &&|p,\bar{p}\rangle,~~(p\neq 0),
  \Label{hwsWinfp} \\
  &&|n\rangle \eqdef
  \left \{
  \begin{array}{ll}
    (\alpha_{-1})^n |0,0\rangle & n\geq 1\\
    |0,0\rangle & n=0  \\
    (\bar{\alpha}_{-1})^{-n} |0,0\rangle & n\leq -1.
  \end{array}
  \right.
  \Label{hwsWinfn}
\eeqa
We can check that these states are the HWS's of $W_{\infty}$.
Their conformal weights are
\beq
  h_{p\bar{p}}=|p|^2,~~h_n=|n|.
\eeq

Neglecting the dependence on the eigenvalues of higher-spin generators,
we consider the characters which count conformal weight only,
\beq
  ch^{W_{\infty}}(\tau)
  \eqdef
  \mbox{tr}q^{\tilde{V}^0_0-\frac{2}{24}}.
\eeq
Since $\tilde{V}^i_n$ contains the terms $\bar{\alpha}_0\alpha_n$ and
$\bar{\alpha}_n\alpha_0$, and the momentum $p$ is non-zero for the
continuous series, the set of generators $\tilde{V}^i_n$ is identified
with the set of oscillators $\alpha_n$, $\bar{\alpha}_n$. Therefore the
character formula of the continuous series of $W_{\infty}$ is
\beq
  ch^{W_{\infty}}_{p\bar{p}}(\tau)
  =
  \frac{q^{|p|^2}}{\eta(\tau)^2}.
\eeq
This result was first derived by Bakas and Kiritsis, using the
$Z_{\infty}$ parafermion \cite{BK3}.

For the discrete series, $\bar{\alpha}_0\alpha_n$ and
$\bar{\alpha}_n\alpha_0$ are acting on the state as $0$. So, the number
of the states of the discrete series is less than one of the continuous
series. Let us define the quantum number $B$ as (number of oscillators
without $\bar{}$ ) $-$ (number of oscillators with $\bar{}$ ).
Then $B|n\rangle=n|n\rangle$, and generators of $W_{\infty}$ do not
change $B$ on $|n\rangle$.
By taking the appropriate linear combinations of $\tilde{V}^i_n$, we can
obtain all the oscillators of the form $\bar{\alpha}_n\alpha_m$.
{}From these two facts, the generating function of the characters of
the discrete series is
\beq
  \sum_{n \in \sym{Z}}t^n ch^{W_{\infty}}_n(\tau)
  =
  \mbox{tr}_{Fock}q^{\tilde{V}^0_0-\frac{2}{24}}t^B
  =
  \frac{q^{-\frac{2}{24}}}{\prod_{n=1}^{\infty}(1-tq^n)(1-t^{-1}q^n)}.
\eeq
{}From this, the character formula of the discrete series of
$W_{\infty}$ is expressed as\footnote{
 Recently Bakas and Kiritsis have constructed the non-linear deformation
 of $W_{\infty}$ based on the $SL(2,\symR)_k/U(1)$ coset model,
 and investigated its characters, which include the characters
 of $W_{\infty}$ in the large $k$ limit \cite{BK4}.}
\beqa
  ch^{W_{\infty}}_n(\tau)
  &=&
  \frac{1}{2\eta(\tau)^2}\sum_{m\in \sym{Z}}
   sign(m)(-1)^m q^{mn-\frac{1}{8}}
   ( q^{\frac{1}{2}(m+\frac{1}{2})^2}
    -q^{\frac{1}{2}(m-\frac{1}{2})^2}) \n
  &=&
  \frac{1}{\eta(\tau)^2}\sum_{m=1}^{\infty}
   (-1)^m q^{\frac{1}{2}m(m-1)+mn}(q^m-1).
\eeqa
The relation between continuous and discrete series is
\beq
  \lim_{p\rightarrow 0}ch^{W_{\infty}}_{p\bar{p}}(\tau)
  =
  \sum_{n\in\sym{Z}}ch^{W_{\infty}}_n(\tau).
\eeq

Next we consider the representations of $W_{\infty}^M$ with
$\tilde{c}=2M$ realized by $M$ complex free bosons.
Since the argument is the same as $W_{\infty}$ case, we present
the results only. The HWS's of $W_{\infty}^M$ are
\beqa
  &&|\vec{p},\vec{\bar{p}}\rangle,~~
  \vec{p}=(0,\cdots,0,p_a,0,\cdots,0),~~p_a\neq0,
  \Label{hwsWinfMp} \\
  &&|n\rangle \eqdef
  \left \{
  \begin{array}{ll}
    (\alpha^1_{-1})^n |\vec{0},\vec{0}\rangle & n\geq 1\\
    |\vec{0},\vec{0}\rangle & n=0  \\
    (\bar{\alpha}^1_{-1})^{-n} |\vec{0},\vec{0}\rangle & n\leq -1.
  \end{array}
  \right.
  \Label{hwsWinfMn}
\eeqa
The degeneracy of the ground states are $1$ and ${M+|n|-1\choose |n|}$
respectively. The conformal weights are
\beq
  h_{\vec{p}\vec{\bar{p}}}=|\vec{p}|^2,~~h_n=|n|.
\eeq

The characters which count conformal weight only, are defined by
\beq
  ch^{W_{\infty}^M}(\tau)
  \eqdef
  \mbox{tr}q^{\tilde{V}^0_0-\frac{2M}{24}}.
\eeq
The character formula of the continuous series of $W_{\infty}^M$ is
\beq
  ch^{W_{\infty}^M}_{\vec{p}\vec{\bar{p}}}(\tau)
  =
  \frac{q^{|\vec{p}|^2}}{\eta(\tau)^{2M}}.
\eeq
The generating function of the character formulas of the discrete series
is
\beq
  \sum_{n \in \sym{Z}}t^n ch^{W_{\infty}^M}_n(\tau)
  =
  \left(
  \frac{q^{-\frac{2}{24}}}{\prod_{n=1}^{\infty}(1-tq^n)(1-t^{-1}q^n)}
  \right)^M.
\eeq
The relation between continuous and discrete series is
\beq
  \lim_{\vec{p}\rightarrow \vec{0}}
  ch^{W_{\infty}^M}_{\vec{p}\vec{\bar{p}}}(\tau)
  =
  \sum_{n\in\sym{Z}}ch^{W_{\infty}^M}_n(\tau).
\eeq

\section {Representations of $W_{\infty}^{1,1}$ and $W_{\infty}^{M,N}$}

We first consider the representations of $W_{\infty}^{1,1}$ with
$\tilde{c}=2$, $c=1$ realized by one pair of complex free boson and
fermion.
Each state in the Fock space is expressed as a linear combination of
$|\tilde{*}\rangle\otimes |*\rangle$, where the first and second factors
are states in the boson and fermion Fock spaces respectively.
$W_{\infty}^{1,1}$ contains $W_{\infty}$ and
$W_{1+\infty}$ as subalgebras, and the generators of $W_{\infty}$ and
$W_{1+\infty}$ are expressed by a boson and a fermion respectively.
Therefore $|\tilde{*}\rangle$ and $|*\rangle$ are the HWS's of
$W_{\infty}$ and $W_{1+\infty}$ respectively. By using the results
obtained in the previous sections, we find the HWS's of
$W_{\infty}^{1,1}$, continuous series and discrete series:
\beqa
  &&|p,\bar{p}\rangle\otimes |0\rangle, \\
  &&|n\rangle \eqdef
  \left \{
  \begin{array}{ll}
    |n-1\rangle\otimes |1\rangle & n\geq 1\\
    |0\rangle\otimes |0\rangle & n=0  \\
    |n+1\rangle\otimes |-1\rangle & n\leq -1.
  \end{array}
  \right.
\eeqa
The first and second factors of the tensor product are
\eql{hwsWinfp}\eqr{hwsWinfn} and \eq{hwsW1inf} respectively.
Their conformal weight $h$ and $U(1)$ charge $Q$ are
\beqa
  &&h_{p\bar{p}}=|p|^2,~~Q_{p\bar{p}}=0, \\
  &&(h_n,Q_n)=
  \left \{
  \begin{array}{ll}
    (n-\frac{1}{2},1) & n\geq 1 \\
    (0,0) & n=0 \\
    (-n-\frac{1}{2},-1) & n\leq -1.
  \end{array}
  \right.
\eeqa

Neglecting the dependence on the eigenvalues of higher-spin generators,
we consider the characters which count conformal weight and $U(1)$
charge only,
\beq
  ch^{W_{\infty}^{1,1}}(\theta,\tau)
  \eqdef
  \mbox{tr}q^{\tilde{V}^0_0-\frac{2}{24}+V^0_0-\frac{1}{24}}
           e^{i\theta J_0}.
\eeq
$\tilde{V}^i_n$ contains the terms $\bar{\alpha}_0\alpha_n$ and
$\bar{\alpha}_n\alpha_0$, and $G^i_n$ and $\bar{G}^i_n$ contain the
terms $\bar{\alpha}_0\psi_n$ and $\alpha_0\bar{\psi}_n$. For the
continuous series, the momentum $p$ is non-zero, so the set of
generators of $W_{\infty}^{1,1}$ is identified with the set of
oscillators $\psi_n$, $\bar{\psi}_n$, $\alpha_n$, $\bar{\alpha}_n$.
Therefore the character of the continuous series is
\beqa
  ch^{W_{\infty}^{1,1}}_{p\bar{p}}(\theta,\tau)
  &\!\!=\!\!&
  \frac{q^{|p|^2}}{\eta(\tau)^2}
  q^{-\frac{1}{24}}
  \prod_{n=1}^{\infty}
  (1+zq^{n-\frac{1}{2}})(1+z^{-1}q^{n-\frac{1}{2}}) \\
  &\!\!=\!\!&
  ch^{W_{\infty}}_{p\bar{p}}(\tau)
  \sum_{n\in\sym{Z}}ch^{W_{1+\infty}}_n(\theta,\tau) \\
  &\!\!=\!\!&
  ch^{W_{\infty}}_{p\bar{p}}(\tau)
  f_{1,0}(\theta,\tau),
\eeqa
where we define $f_{K,Q}(\theta,\tau)$ as
\beq
  f_{K,Q}(\theta,\tau)
  \eqdef
  \frac{1}{\eta(\tau)} \sum_{n\in\sym{Z}}
  q^{\frac{K}{2}(n+\frac{Q}{K})^2}z^{K(n+\frac{Q}{K})}.
\eeq

For the discrete series, $B|n\rangle=n|n\rangle$, and the generators of
$W_{\infty}^{1,1}$ do not change $B$ on $|n\rangle$. By taking
appropriate linear combinations of generators of $W_{\infty}^{1,1}$,
we obtain all the oscillators of the form
$\bar{\psi}_n\psi_m$, $\bar{\alpha}_n\psi_m$,
$\alpha_n\bar{\psi}_m$, $\bar{\alpha}_n\alpha_m$.
{}From these two facts, the generating function of the characters of
the discrete series is
\beq
  \sum_{n \in \sym{Z}}t^n ch^{W_{\infty}^{1,1}}_n(\theta,\tau)
  =
  \frac{q^{-\frac{2}{24}}}{\prod_{n=1}^{\infty}(1-tq^n)(1-t^{-1}q^n)}
  \cdot
  q^{-\frac{1}{24}}
  \prod_{n=1}^{\infty}
  (1+tzq^{n-\frac{1}{2}})(1+t^{-1}z^{-1}q^{n-\frac{1}{2}}).
\eeq
{}From this equation, the character formula of the discrete series is
\beq
  ch^{W_{\infty}^{1,1}}_n(\theta,\tau)
  =
  \sum_{\ell\in\sym{Z}}
  ch^{W_{\infty}}_{n-\ell}(\tau)
  ch^{W_{1+\infty}}_{\ell}(\theta,\tau).
\eeq
The relation between continuous and discrete series is
\beq
  \lim_{p\rightarrow 0}
  ch^{W_{\infty}^{1,1}}_{p\bar{p}}(\theta,\tau)
  =
  \sum_{n\in\sym{Z}}ch^{W_{\infty}^{1,1}}_n(\theta,\tau).
\eeq

Since $W_{\infty}^{1,1}$ is a superalgebra, we must consider the $R$
sector also. By using the spectral flow, the character of the $R$ sector
is expressed in terms of the character of the $NS$ sector:
\beq
  ch^{W_{\infty}^{1,1},R}(\theta,\tau)
  =
  q^{\frac{3}{24}}z^{\frac{3}{6}}
  ch^{W_{\infty}^{1,1}}(\theta+\pi\tau,\tau).
\eeq
Explicitly they are
\beqa
  ch^{W_{\infty}^{1,1},R}_{p\bar{p}}(\theta,\tau)
  &\!\!=\!\!&
  ch^{W_{\infty}}_{p\bar{p}}(\tau)
  f_{1,\frac{1}{2}}(\theta,\tau), \\
  ch^{W_{\infty}^{1,1},R}_n(\theta,\tau)
  &\!\!=\!\!&
  \sum_{\ell \in \sym{Z}}
  ch^{W_{\infty}}_{n-\ell}(\tau)
  \chi^{\hat{u}(1)_1}_{\ell+\frac{1}{2}}(\theta,\tau).
\eeqa
The conformal weight $h^R$ and $U(1)$ charge $Q^R$ in the $R$ sector are
\beqa
  &&h_{p\bar{p}}^R=\frac{1}{8}+|p|^2,~~Q_{p\bar{p}}^R=\frac{1}{2} \\
  &&(h_n^R,Q_n^R)=
  \left \{
  \begin{array}{ll}
    (\frac{1}{8}+n,\frac{3}{2}) & n\geq 1 \\
    (\frac{1}{8},\frac{1}{2}) & n=0 \\
    (\frac{1}{8}-(n+1),-\frac{1}{2}) & n\leq -1.
  \end{array}
  \right.
\eeqa
The ground states are singlets for $n=0,-1$, and doublets for others.

In order to study whether a supersymmetry exist or not, and if it
exists, whether it is broken or unbroken, we define the Witten index:
\beq
  Index \eqdef
  \mbox{tr}_R
  q^{\tilde{V}^0_0+V^0_0-h^R}(-1)^F,
  \Label{index}
\eeq
where $F$ is the fermion number, and trace is taken over the
representation space of the $R$ sector.
By using the property of the spectral flow and the fact that the fermion
numbers of the generators of $W_{\infty}^{1,1}$ agree with their $U(1)$
charges, the Witten index is expressed as follows:
\beq
  Index
  =
  q^{\frac{3}{24}-h-\frac{1}{2}Q}
  ch^{W_{\infty}^{1,1}}(\pi+\pi\tau,\tau).
\eeq
For representations $n=0,-1$, the Witten indices are
\beq
  Index_0=1,~~Index_{-1}=-1.
\eeq
For other representations, the Witten index vanishes. Therefore there
exists a ($N=2$) supersymmetry for all representations and it is
broken for $n=0,-1$.

Next we consider the representations of $W_{\infty}^{M,N}$ with
$\tilde{c}=2M$, $c=N$ realized by $M$ complex free bosons and $N$
complex free fermions. Since the argument is the same as
$W_{\infty}^{1,1}$ case, we present the results only. The HWS's
of $W_{\infty}^{M,N}$ are
\beqa
  &&|\vec{p},\vec{\bar{p}}\rangle\otimes |0\rangle, \\
  &&|n\rangle \eqdef
  \left \{
  \begin{array}{ll}
    |n-N\rangle\otimes |N\rangle & n\geq N\\
    |0\rangle\otimes |n\rangle & -N<n<N  \\
    |n+N\rangle\otimes |-N\rangle & n\leq -N.
  \end{array}
  \right.
\eeqa
The first and second factors of the tensor product are given by
\eql{hwsWinfMp}\eqr{hwsWinfMn} and \eq{hwsW1infN}
respectively. Their conformal weight $h$ and $U(1)$ charge $Q$ are
\beqa
  &&h_{\vec{p}\vec{\bar{p}}}=|\vec{p}|^2,~~Q_{\vec{p}\vec{\bar{p}}}=0, \\
  &&(h_n,Q_n)=
  \left \{
  \begin{array}{ll}
    (n-\frac{1}{2}N,N) & n\geq N \\
    (\frac{1}{2}|n|,n) & -N<n<N \\
    (-n-\frac{1}{2}N,-N) & n\leq -N.
  \end{array}
  \right.
\eeqa

Neglecting the dependence on the eigenvalues of higher-spin generators,
we consider the characters which count conformal weight, $U(1)$
charge and eigenvalues of $SU(N)$,
\beq
  ch^{W_{\infty}^{M,N}}(\theta,\vec{\theta},\tau)
  \eqdef
  \mbox{tr}q^{\tilde{V}^0_0-\frac{2M}{24}+V^0_0-\frac{N}{24}}
           e^{i\theta J_0}e^{i\vec{\theta}\cdot\vec{H}_0}.
\eeq
The character formula of the continuous series is
\beqa
  ch^{W_{\infty}^{M,N}}_{\vec{p}\vec{\bar{p}}}(\theta,\vec{\theta},\tau)
  &\!\!=\!\!&
  \frac{q^{|\vec{p}|^2}}{\eta(\tau)^{2M}}
  q^{-\frac{N}{24}}
  \prod_{j=1}^{N} \prod_{n=1}^{\infty}
  (1+zz_{j-1}^{-1}z_jq^{n-\frac{1}{2}})
  (1+z^{-1}z_{j-1}z_j^{-1}q^{n-\frac{1}{2}}) \\
  &\!\!=\!\!&
  ch^{W_{\infty}^M}_{\vec{p}\vec{\bar{p}}}(\tau)
  \sum_{n\in\sym{Z}}
  ch^{W_{1+\infty}^N}_n(\theta,\vec{\theta},\tau) \\
  &\!\!=\!\!&
  ch^{W_{\infty}^M}_{\vec{p}\vec{\bar{p}}}(\tau)
  \sum_{a=0}^{N-1}
  f_{N,a}(\theta,\tau)
  \chi^{\widehat{su}(N)_1}_a(\vec{\theta},\tau).
\eeqa
The generating function of the characters of the discrete series is
\beqa
  \hspace{-4mm}
  \sum_{n \in \sym{Z}}t^n
  ch^{W_{\infty}^{M,N}}_n(\theta,\vec{\theta},\tau)
  &\!\!=\!\!&
  \left(
  \frac{q^{-\frac{2}{24}}}{\prod_{n=1}^{\infty}(1-tq^n)(1-t^{-1}q^n)}
  \right)^M \n
  &&
  \times
  q^{-\frac{N}{24}}
  \prod_{j=1}^{N} \prod_{n=1}^{\infty}
  (1+tzz_{j-1}^{-1}z_jq^{n-\frac{1}{2}})
  (1+t^{-1}z^{-1}z_{j-1}z_j^{-1}q^{n-\frac{1}{2}}).
\eeqa
{}From this, the character formula of the discrete series is
\beq
  ch^{W_{\infty}^{M,N}}_n(\theta,\vec{\theta},\tau)
  =
  \sum_{\ell \in \sym{Z}}
  ch^{W_{\infty}^M}_{n-\ell}(\tau)
  ch^{W_{1+\infty}^N}_{\ell}(\theta,\vec{\theta},\tau).
\eeq
The relation between continuous and discrete series is
\beq
  \lim_{\vec{p}\rightarrow \vec{0}}
  ch^{W_{\infty}^{M,N}}_{\vec{p}\vec{\bar{p}}}(\theta,\vec{\theta},\tau)
  =
  \sum_{n\in\sym{Z}}ch^{W_{\infty}^{M,N}}_n(\theta,\vec{\theta},\tau).
\eeq

By using the spectral flow, the character of the $R$ sector is expressed
in terms of the character of the $NS$ sector:
\beq
  ch^{W_{\infty}^{M,N},R}(\theta,\vec{\theta},\tau)
  =
  q^{\frac{3N}{24}}z^{\frac{3N}{6}}
  ch^{W_{\infty}^{M,N}}(\theta+\pi\tau,\vec{\theta},\tau).
\eeq
Explicitly they are
\beqa
  ch^{W_{\infty}^{M,N},R}_{\vec{p},\vec{\bar{p}}}(\theta,\vec{\theta},\tau)
  &\!\!=\!\!&
  ch^{W_{\infty}^M}_{\vec{p},\vec{\bar{p}}}(\tau)
  \sum_{a=0}^{N-1}
  f_{N,a+\frac{N}{2}}(\theta,\tau)
  \chi^{\widehat{su}(N)_1}_a(\vec{\theta},\tau) \\
  ch^{W_{\infty}^{M,N},R}_n(\theta,\vec{\theta},\tau)
  &\!\!=\!\!&
  \sum_{\ell \in \sym{Z}}
  ch^{W^M_{\infty}}_{n-\ell}(\tau)
  \chi^{\hat{u}(1)_N}_{\ell+\frac{N}{2}}(\theta,\tau)
  \chi^{\widehat{su}(N)_1}_a(\vec{\theta},\tau),
\eeqa
where, in the second equation, $a\equiv \ell$ $(\mbox{mod }N)$.
The conformal weight $h^R$, $U(1)$ charge $Q^R$ and the degeneracy
of the ground states in the $R$ sector are
\beqa
  &&
  h_{\vec{p}\vec{\bar{p}}}^R=\frac{1}{8}N+|\vec{p}|^2,
  ~~Q_{\vec{p}\vec{\bar{p}}}^R=\frac{1}{2}N,~~
  \mbox{degeneracy}=\sum_{a=0}^N{N\choose a}=2^N, \\
  &&
  (h_n^R,Q_n^R,\mbox{degeneracy}) \n
  &&=
  \left \{
  \begin{array}{ll}
    (\frac{1}{8}N+n,\frac{3}{2}N,
     \sum_{a=0}^N{N\choose a}{M+n-a-1\choose n-a}) & n\geq N \\
    (\frac{1}{8}N+n,\frac{1}{2}N+n,
     \sum_{a=0}^n{N\choose a}{M+n-a-1\choose n-a}) & 0<n<N \\
    (\frac{1}{8}N,\frac{1}{2}N+n,
     {N\choose N+n}) & -N\leq n \leq 0 \\
    (\frac{1}{8}N-(N+n),-\frac{1}{2}N,
     \sum_{a=2N+n}^N{N\choose a}{M-n-2N+a-1\choose -n-2N+a}) & -2N<n<-N \\
    (\frac{1}{8}N-(N+n),-\frac{1}{2}N,
     \sum_{a=0}^N{N\choose a}{M-n-2N+a-1\choose -n-2N+a}) & n\leq -2N.
  \end{array}
  \right.
\eeqa

By using the property of the spectral flow and the fact that the fermion
numbers of the generators of $W_{\infty}^{M,N}$ agree with their $U(1)$
charges, the Witten index \eq{index} is expressed as
\beq
  Index=
  q^{\frac{N+2M}{24}-h-\frac{1}{2}Q}
  ch^{W_{\infty}^{M,N}}(\pi+\pi\tau,\vec{0},\tau).
\eeq
For the continuous series, the Witten index vanishes for all $M,N$.
Therefore, for the continuous series, there exists a supersymmetry,
and it is unbroken.

In the case of the discrete series with $M\neq N$, the Witten index is
not a number but a $q$-series. Namely, at excited state, the number of
bosonic states does not agree with one of fermionic states. Therefore a
supersymmetry does not exist in the discrete series of
$W_{\infty}^{M,N}$ $(M\neq N)$.
In the case of the discrete series with $M=N$, the Witten index is just
a number. So a ($2N^2$ extended) supersymmetry exists. For
representations $n$ $(-N\leq n \leq 0)$, the Witten index is
\beq
  Index_n=(-1)^n{N\choose N+n},
\eeq
and a supersymmetry is broken. For other representations, the Witten
index vanishes and a supersymmetry is unbroken.

\section {Discussion}
In this paper we have studied the irreducible unitary highest weight
representations of $W$ infinity algebras, which are obtained from free
field realizations, and derived their character formulas. We have also
constructed a new superalgebra $W_{\infty}^{M,N}$, whose bosonic sector
is $W_{\infty}^M\oplus W_{1+\infty}^N$. Its representations obtained
from a free field realization are classified into two classes,
continuous and discrete. There exists a supersymmetry in the
continuous series, whereas a supersymmetry exists only for $M=N$ in
the discrete series. This is expected from the
counting of the bosonic and fermionic degrees of freedom of the
generators:
\beq
  \begin{array}{ccccccccc}
    W^{i-1,(\alpha\beta)}(z)&&
    \tilde{W}^{i,(ab)}(z)&&
    G^{i,a\alpha}(z)&&
    \bar{G}^{i,a\alpha}(z)&&\\
    N^2&\!\!\!\!+\!\!\!\!&M^2&\!\!\!\!-\!\!\!\!&
      MN&\!\!\!\!-\!\!\!\!&MN&\!\!=&(M-N)^2.
  \end{array}
\eeq
Perhaps a supersymmetry in the continuous series for $M\neq N$ may be
an accidental one.

The representations with higher central charge and the realization
independent representations are future subjects. There are two
difficulties in developing the realization independent representation
theories of $W$ infinity algebras. One is that there are infinite
number of fields. The other is the complicated dependence on the
eigenvalues of higher-spin generators. For example, no one has succeeded
in computing even the level $1$ Kac determinant.

Although our representation theory is a restricted one, we have obtained
the character formulas. It is interesting to apply these character
formulas to models with $W$ infinity symmetry, for example \cite{BK2},
integrable non-linear differential equation systems
such as the $KP$ hierarchy and the Toda hierarchy,
four dimensional self-dual gravity \cite{Park}, Virasoro ($W$)
constraints on the partition function of the multi-matrix model
\cite{FKNDVV}, and $W$ infinity gravity \cite{wG,SSN}.
The modular properties of the characters are also subjects of
future research.

Finally, we mention the anomaly-free conditions.
In refs.\cite{Ya,PRSgh}, the anomaly-free conditions for $W_{\infty}$,
$W_{1+\infty}$ and $W_{\infty}^{1,1}$ are considered by the BRS
formalism and $\zeta$ function regularization, and it is shown that
their critical central charges are $-2$, $0$ and $-3$ respectively.
Similar calculations have been done for $W_{1+\infty}^N$ and
$W_{\infty}^{1,N}$ by considering the ghost realizations of them,
and their critical central charges are $0$ and $-2-N$ respectively
\cite{OS2}. For $W_{\infty}^{M,N}$, we obtain $k_{ghost}=M$, and
the critical central charge is
\beq
  (\tilde{c}+c)_{critical}=-(\tilde{c}+c)_{ghost}=-2M^2-MN.
\eeq
%
%
\section*{Acknowledgments}
\vspace{-3mm}
\noindent
The author would like to thank T. Inami for
discussions and comments on the manuscript.
He would like to acknowledge useful discussions with
H. Nohara and T. Sano.
\newpage

%
%
\end{document}